# How Sovereign Is Sovereign Compute? A Review of 775 Non-U.S. Data Centers

Aris Richardson
RAND
arislrichardson@gmail.com

Haley Yi
Independent

Michelle Nie
The Center for a New American Security
michellesnie@gmail.com

Simon Wisdom
Independent

Casey Price
Independent

Ruben Weijers
Delft University of Technology

Steven Veld
The Institute of AI Policy and Strategy

Mauricio Baker
RAND

***Abstract***

Previous literature has proposed that the companies operating data centers enforce government regulations on AI companies. Using a new dataset of 775 non-U.S. data center projects, this paper estimates how often data centers could be subject to foreign legal authorities due to the nationality of the data center operators. We find that U.S. companies operate 48% of all non-U.S. data center projects in our dataset when weighted by investment value—a proxy for compute capacity. This is an approximation based on public data and should be interpreted as an initial estimate. For the United States, our findings suggest that data center operators offer a lever for internationally governing AI that complements traditional export controls, since operators can be used to regulate computing resources already deployed in non-U.S. data centers. For other countries, our results show that building data centers locally does not guarantee digital sovereignty if those facilities are run by foreign entities. To support future research, we release our dataset, which documents over 20 variables relating to each data center, including the year it was announced, the investment value, and its operator's national affiliation.[1] The dataset also includes over 1,000 quotes describing these data centers' strategic motivations, operational challenges, and engagement with U.S. and Chinese entities.

## 1 Introduction

Previous literature has proposed that the companies operating data center hardware could serve as key intermediaries between governments and AI companies that use data centers for development and deployment of AI (Heim et al. 2024). Data centers are maintained by operators, which are not always headquartered in the country in which the data center is located. This opens up the potential for governments to use domestically headquartered operators as jurisdictional hooks for data center activities abroad. Jurisdictional hooks are a legal basis for a government to claim jurisdiction over matters that would otherwise be outside its authority.

[1] The dataset can be found at https://github.com/alarichardson/non-us-data-center-registry.

Under international law, as reflected in the *Restatement (Fourth) of Foreign Relations Law of the United States*, a state may claim jurisdiction to something based on several recognized principles. Among them, the *territorial principle* is commonly evoked for data centers. The territorial principle allows a state to exercise jurisdiction over conduct that occurs within its territory or has substantial effects there. Due to the principle of territoriality, data centers are often subject to jurisdictional claims from the country in which they are physically located.

A large share of the world's data centers has been built on U.S. soil, giving the U.S. government jurisdictional claims to most of the world's compute capacity. Compute capacity refers to the ability of a system, typically a data center, to perform computational tasks as measured in floating point operations (FLOP). As Sevilla et al. (2022) show, frontier AI models require vast amounts of compute, making access to this capacity essential for AI development and deployment. Analysis by the Synergy Research Group indicates that the United States still accounts for an estimated 54% of worldwide compute capacity (Synergy Research Group 2025).

Many countries have expressed concern that relying on data centers based in another country undermines their sovereignty. Sovereignty traditionally refers to a state's supreme authority within its territorial boundaries (Philpott 2024). Since the early 2000s, policymakers have extended the concept into the digital realm, arguing that states have the right to control their own digital environments. The term digital sovereignty appears frequently in government statements around the world and is typically used to motivate requirements for infrastructure to be built or stored within a state's borders (Pohle and Thiel 2020). For instance, India's Data Center Policy emphasizes the need to host data centers domestically to protect national interests, and the European Commission's AI Continent Action Plan frames domestic data centers and clouds as essential for European sovereignty (Ministry of Electronics and Information Technology 2020; "The AI Continent Action Plan" 2025).

What remains unclear, however, is how much these efforts to localize data centers actually translate into meaningful shifts in jurisdictional control. When a government claims jurisdiction over a data center, that government does not necessarily gain control over that infrastructure or data. Multiple governments may assert overlapping claims—including the governments of the country where the data center is located, the country where the operating company is headquartered, or the country of the users whose data is stored there. Whether any one government's claim results in actual control—such as the ability to compel access to data, restrict use of compute, or enforce compliance—depends on enforcement power, company cooperation, and existing international agreements.

Although it is less commonly evoked than territoriality, the *nationality principle* is another recognized principle that allows a state to assert jurisdiction over its nationals—both individuals and legal entities such as corporations—even when their conduct occurs abroad. Under the principle of nationality, data center operators can serve as jurisdictional "hooks" into data centers abroad. These operators could function as record-keepers, verifiers of AI companies' commitments related to compute use, and enforcers of regulations (Heim et al. 2024).

The resources within data centers can either be operated by end customers or through an intermediary, such as an Infrastructure as a Service (IaaS) provider. The IaaS market is highly concentrated in U.S. companies Amazon Web Services, Microsoft Azure, and Google Cloud, followed by Chinese providers Alibaba and Tencent (Musin 2021).

Several countries, including the U.S. and China, have passed laws compelling their companies to comply with government data access requests regardless of where the data is physically stored based on the nationality principle and protective principle (U.S. Department of Justice 2023; Covington 2015; CNA 2023). As a result, the U.S. and China may have jurisdictional hooks into a significant portion of the world's data center infrastructure—not only within their own borders, but globally through their IaaS providers.

While prior work has made substantial progress in mapping global AI infrastructure, it often equates the physical location of a data center with that country's control without fully addressing the legal complexities of this assumption. Lehdonvirta et al. (2024) map the global distribution of public cloud GPU deployments to assess national AI capacity, explicitly framing their analysis in terms of territoriality but without engaging other jurisdictional bases. Pilz et al. (2025) curate a global dataset of AI supercomputers, analyzing trends in performance, power, and ownership. While they acknowledge that ownership and jurisdiction may diverge, and that companies may be headquartered in different countries than their infrastructure, they focus on territorial presence in their national comparisons and do not investigate alternative legal hooks in depth. Westgarth et al. (2024) benchmark national compute capacity using indicators such as server installations, energy access, and infrastructure investment. Though they do not explicitly state a territorial premise, their methodology presumes that infrastructure located within national borders implies domestic control.

This paper has two primary contributions. First, it presents a dataset of non-U.S. data center projects that captures both the host country and the nationality of the operator when available. This dataset also documents additional attributes of data center projects (e.g., investment patterns, ownership, public-private partnerships). Second, the paper roughly estimates the percentage of data centers outside the U.S. that other countries, including the U.S., could claim jurisdiction to through data center operators.

Additional variables in the dataset help interpret the impact of certain actors having jurisdictional hooks into particular types of infrastructure. For example, a country may have many data centers, but if its most expensive, AI-dedicated data centers are operated by U.S. companies, its AI infrastructure may effectively fall under foreign jurisdiction. This dataset also provides a baseline for future research: for example, if one country experiences more delays or cancellations than others, researchers can investigate whether those setbacks stem from foreign partners or domestic conditions.

All data centers in our sample are located outside of the United States because much of the world's compute infrastructure is already concentrated within U.S. borders (Pilz and Heim 2023)—and is thus clearly under U.S. jurisdiction. By examining infrastructure built elsewhere, often as part of efforts to bring compute under national control, researchers can better assess the extent to which digital sovereignty efforts reduce foreign control over domestic compute.

The findings from our estimates and accompanying data have direct implications for U.S. policymakers. They offer a clearer picture of how often data centers abroad could fall under U.S. jurisdiction and, if not, to whom the jurisdiction could fall instead. This information can support strategy by helping the U.S. prioritize countries to engage when enforcing AI strategy internationally.

For non-U.S. countries, the findings have different but equally important implications. Governments that have invested in localized infrastructure under the banner of digital sovereignty must grapple with the possibility that these efforts may not deliver meaningful jurisdictional control if the most strategically important facilities—particularly those built for AI—remain operated by foreign firms. In these cases, alignment with foreign regulatory regimes, including those of the U.S., may be less a matter of choice than a result of structural reliance on foreign compute providers. Countries facing such constraints may choose to redirect their investments toward more targeted goals, such as AI applications or workforce development, or cooperate with foreign firms to shape governance outcomes more directly.

The dataset collected for this paper is designed to support future policy research. It includes over 20 variables for each data center, such as investment value, year announced, and whether projects experienced delays or cancellations. These allow researchers to investigate other questions about

non-U.S. data centers. The dataset also includes over 1,000 quotes that capture how non-U.S. actors describe their motivations, challenges, and their opinions on involvement of U.S. and Chinese firms. Research on this data can be used to align future compute and AI investments with policymakers' priorities.

## 2 Methodology

To develop our estimates, we compile open-source information to develop a sample of data centers globally. For each data center, we document the national origin of their operators, funders, and construction partners. We also collect over 20 variables on each data center, including year announced, investment value, delays, cancellations, and involvement of foreign companies to contextualize jurisdictional influence.

### 2.1 Data collection

This dataset catalogs 775 data center projects across 123 countries (after searching 193 countries), with data collection ending in the third quarter of 2024. These data center projects include existing and planned initiatives to build data centers, supercomputers, and compute clouds. To standardize terminology, we refer to all initiatives as "data center projects"—even when sources use other terms or when a single project establishes more than one compute cluster or facility. We deliberately exclude data on data centers in the U.S. in order to focus our analysis on U.S. influence abroad.

The data in our dataset was collected from a mix of public sources: news articles, government announcements, and industry reports. The data search process began with individual queries for each of the 193 countries using the AI-augmented search tool Perplexity AI in Q2 and Q3 of 2024. Researchers used a standardized prompt to search Perplexity for each country: "Does [country] have any sovereign compute projects, such as data centers, supercomputers, sovereign clouds, or partnerships with large companies to build a sovereign compute project?"

Like Google Search, Perplexity returns the top-ranked internet results relevant to a query. In contrast to a typical search engine, Perplexity had two strengths:

1. A single prompt can retrieve information that may be labeled with different terms in public press (e.g., supercomputers, data centers, and cloud infrastructure), reducing the need for multiple keyword variations.
2. When asked whether a country has any sovereign compute projects, Perplexity can explicitly return "no" rather than returning unrelated or tangential results.

No information was recorded directly from AI-generated summaries; all data entered into our dataset was reviewed manually. If Perplexity returned no clear results for a country, we conducted additional searches manually using Google Search and industry databases such as Data Center Dynamics with predetermined key terms.

As individual data centers were identified, researchers recorded 26 variables for each project along with the URLs of the sources of our information. These URLs were then processed using a Python script to extract additional information on the same variables. All extracted data was manually reviewed before it was included in the dataset.

To ensure consistency across entries, the researchers were given feedback from a team lead weekly during data collection and met weekly during the data collection phase to clarify questions and reduce discrepancies.

### 2.2 Data cleaning

The data was then cleaned and standardized to allow for comparison across projects (e.g., converting all investment figures to USD, not adjusted for inflation), resulting in the set of 775 projects analyzed in this paper.

Some collected data unrelated to data centers were cleaned and retained in a separate section of the dataset for future researchers to use but are not included in the analysis presented here. Such information includes national digital strategies, investments in local workforce training, and investments in local semiconductor manufacturing.

We marked a company as an operator if it either ran the data center's operations or provided cloud services. We started with a list of well-known IaaS companies and labeled those companies as operators by default:

- U.S.-headquartered: AWS, Microsoft, Google, Oracle, IBM, Equinix, Dell, VMware, Raxio Group/Roha Group, Gennext Technologies
- China-headquartered: Huawei, Alibaba, Baidu, China Mobile, China Telecom, China Unicom, Lenovo, ZTE, Inspur, Sugon, 21Vianet, ByteDance
- Headquartered in other countries: Paratus, Orange, Telefónica, OVH

We manually reviewed each instance of these companies to check that the company was actually acting as an operator. If a company was only providing financing or hardware, it was not counted as an IaaS provider but its presence was still recorded for frequency analysis.

### 2.3 Dataset variables

Information about each data center project was coded using a standardized set of variables to enable structured comparisons across countries, ownership models, and types of infrastructure. Core metadata include the country, year of announcement, reported project type (e.g., data center, cloud, supercomputer), and ownership model (public, private, public-private). We also record the intended use case (e.g., government, healthcare, R&D), reported setbacks, and whether the project reportedly involves AI/ML, edge computing, or cybersecurity applications.

To enable analysis of jurisdictional levers, we documented each instance of actor involvement by both sector and nationality. These included U.S. government involvement, U.S. company involvement, Chinese government involvement, Chinese company involvement, domestic government and company involvement, and other international actors (government or corporate).

When available, quantitative variables such as investments ($) and number of GPUs in the projects were recorded to contextualize the compute capacity of the data center projects. While exact metrics like number of GPUs or FLOPs were not public for most projects, we use reported investment value in the data center projects as a proxy for compute capacity.

We also recorded quotes from the articles to compile a qualitative portion of our dataset, which includes descriptions of project motivations, U.S. and Chinese involvement, and the specific types of setbacks encountered by projects. Setbacks were briefly labeled by the type of problem encountered and checked for frequency by country, but otherwise this qualitative data is not analyzed in this paper.

### 2.4 Analysis

We produced estimates of how frequently certain actors—especially U.S. and Chinese companies—are operators or play other key roles (sponsors, hardware providers, construction partners) for non-U.S. data center projects.

Our estimates used two types of measures: unweighted frequency and investment-weighted frequency. By weighing data centers by their investment value, our analysis accounted for differences in the sizes of data centers.

In addition to producing estimates, we mapped the geographic distribution of key actors and compared the global presence of U.S. and Chinese operators.

### 2.5 Limitations

While the dataset provides a detailed overview of documented projects, the quality and quantity of the dataset are limited by the availability of publicly reported information at the time of data collection. Our final 775 data centers are a sample of the total data centers announced in recent years. Some nations deliberately underreport their computing infrastructure activities, especially for sensitive projects like supercomputing, leading to gaps in representation, especially on projects located in China. Due to the scarcity of public information, many entries lack precise financial or technical reporting. Furthermore, many of the variables reflect the language used in original sources, which is not standardized across sources (e.g., the type of compute cluster created, whether a project is statedly built for AI applications).

Our rough estimate assumes data center investments are proportional to the data center's compute capacity, so this estimate does not account for any investment costs that scale non-linearly (e.g., discounts for bulk purchases) or which have different costs across countries (e.g., construction labor, internationally shipped equipment).

## 3 Results

### 3.1 Applicability

Our dataset accounts for 360 billion dollars of existing and planned investments in data centers. See Figure 4 (in Appendix 4). Our data is most applicable to data centers announced between 2020 and 2024 across the private and public sectors.

#### 3.1.1 Projects over time

We found an increase in the total number of non-U.S. data center projects per year over time, with a sharp increase in the number of projects announced in 2020. See Figure 1 (in Appendix 1).

### 3.2 Ownership structures

Private, public, and public-private partnerships appeared in similar numbers across the dataset, though project types differ by ownership structure. Nearly all projects reported as supercomputers were public or public-private partnerships, as opposed to private projects. Meanwhile, data centers were most commonly private projects and compute clouds were most frequently public-private partnerships.

### 3.3 Key actors: operators

#### 3.3.1 U.S. operators

U.S. companies served as operators for 18% of data center projects in our dataset, accounting for 48% of total data center value. Among projects that were reported to have AI applications, U.S. companies operated 27% of projects and 56% of value. See Table 1 (in Appendix 2).

#### 3.3.2 Chinese operators

Chinese companies served as operators for 8% of data center projects in our dataset, accounting for 5% of total investment value. Among projects that were reported to have AI applications, Chinese companies operated 6% of projects and 0% of total AI-related investment value (Table 1). This includes firms headquartered in China that operate data centers both inside China and in other countries. The investment value associated with their operations solely outside of China is shown separately in Figure 1.

These figures likely underestimate the global value of compute that Chinese companies operate, due to data availability challenges.

### 3.3.3 Non-U.S. and non-Chinese operators

A significant share of total investment value was not associated with U.S. or Chinese operators: 74% of all data centers, 47% of total value, 65% of AI-designated data centers, and 44% of AI data center value (Table 1). In countries such as India, Taiwan, South Korea, and South Africa, we identify over $10 billion in existing or planned data center projects with no identifiable U.S. or Chinese operator. See Figure 2 (in Appendix 3).

Within this group, domestic operators accounted for 10% of all projects and 11% of value (16% of AI projects; 19% of AI value). Foreign operators not based in the U.S. or China represented 12% of all projects but contributed negligible value (1% of AI projects; 0% of AI value). A significant proportion—53% of all projects and 36% of total value—had no identified operator at all (48% of AI projects; 25% of AI value).

### 3.3.4 Comparing U.S. and Chinese operators

We compared U.S. operators to Chinese operators geographically, weighted by investment value. See Figure 4 (in Appendix 4). This analysis reflects investment value rather than frequency of operators, since frequency can give a skewed impression of the compute capacity U.S. and Chinese companies have jurisdictional exposure to.

While Chinese operators were the sole source of operator-linked investment value in 8 countries, U.S. operators filled that role in 26. U.S. operators were common throughout the Americas, Europe, and select countries in the Middle East. Chinese companies, by contrast, operated in a narrower group of countries, concentrated in Africa (e.g., Ghana, Zambia, Senegal) and South Asia (e.g., Bangladesh).

There was limited overlap between U.S. and Chinese operators; only five countries—Chile, Kenya, Namibia, Norway, and Saudi Arabia—had a mix of U.S. and Chinese operators throughout the country. In these mixed cases, U.S. companies typically accounted for the larger share of investment value, with the exception of Namibia, where investment was evenly split.

## 3.4 Key actors: Any role

We identified times that government and non-government entities play any of the following roles in a data center: operator, fiscal sponsor, hardware supplier/assembler, owner. We found that the majority of data center projects were international collaborations of some kind. Out of 775 entries, we found 537 (68%) were collaborations with another country's government or companies. See Figure 5 (in Appendix 5).

Among government actors, domestic governments were frequent sponsors of the data centers in their country, while foreign governments were sponsors far less often.

The companies that appeared most frequently in the dataset include U.S.-headquartered firms (Microsoft, AWS, Google, Oracle), Chinese firms (Huawei, Alibaba, ZTE), and firms based in France (Orange, OVH), Brazil (Scala), Sweden (Flexenclosure), and Kenya (Africa Data Centers).

### 3.4.1 U.S. companies (any role)

U.S. companies took on a role in 27% of data center projects in our dataset, 68% of total investment value, 41% of AI-designated projects, and 93% of AI project value. Direct U.S. government sponsorship was limited, appearing in just 0.8% of all projects.

### 3.4.2 Chinese companies (any role)

Chinese companies took on a role in 12% of data center projects, 5% of total investment value, 6% of AI-designated projects, and 0.3% of AI project value. State sponsorship was similarly rare, with Chinese government entities involved in just over 1% of projects. Outside of China, Chinese firms had little visible influence over AI-focused infrastructure.

#### 3.4.3 Third-country companies (any role)

Companies not based in the U.S. or China took on a role in the remaining 61% of data center projects, 27% of total investment value, 53% of AI-designated projects, and 7% of AI project value. These projects were often concentrated in regions such as Europe and East Asia. While many were not visibly tied to U.S. firms, there were U.S.-designed GPUs in every AI project with available information on the GPUs in the project. This continued use of U.S. technology highlights the wide-reaching potential of U.S. AI chip export controls.

### 3.5 Setbacks

One prominent trend we noticed was that a significant portion of data centers faced setbacks (12% across the dataset), defined as reports of major barriers such as delays, cancellations, or difficulty securing customers. These were most often attributed to (1) political or regulatory issues in the host country and (2) limited resources—insufficient funding, energy, talent, or semiconductors.

Certain countries experienced disproportionately high rates of disruption. As shown in Table 2 (see Appendix 6), above-average rates of setbacks were not concentrated in any one geographic region; Russia faced the highest frequency (36%), followed by Malaysia, Pakistan, Ireland, the Netherlands, France, Nigeria, Japan, Kenya, and China. The nature of these setbacks varied country to country; some countries, like Russia, faced a broad spectrum of issues, from low resourcing to regulatory constraints, while others, such as the Netherlands, consistently struggled with project planning due to environmental regulations. Notably, countries with the highest prevalence of setbacks do not have the lowest GDP or highest corruption indexes, indicating that localized factors—such as regulatory environments, infrastructure availability, and market conditions—may play a key role in avoiding setbacks. However, this data may be misleading since countries with unstable energy grids or high corruption may be less likely to have local data centers built in the first place.

## 4 Discussion

Although many governments have framed domestic data center construction as a path to digital sovereignty, our dataset reveals that these data centers are often operated by foreign firms. This exposes these data centers to potential jurisdictional hooks from the operators' home governments. In our dataset, foreign companies operated approximately 53% of investment value in non-U.S. data centers. 11% of investment value was confirmed to be associated with domestic operators, indicating that foreign operational control may dominate in practice. As a key limitation to our estimates, 36% of the data center investment value in our dataset was unassigned to an operator.

The U.S. was by far the government with the most potential jurisdictional exposure to non-U.S. compute capacity in our dataset. Using investment value as a rough proxy for compute capacity, our work suggests that the U.S. government could have jurisdictional hooks for approximately 76% of global compute capacity. This is a rough approximation derived from publicly available data, using investment value as a proxy for a data center's compute capacity, and should be interpreted as an initial estimate rather than a definitive measurement. This could be an overestimate, given the lack of public data on Chinese investments in data centers, which might be less likely to have U.S. operators.

To estimate the global share of compute capacity over which the United States may assert jurisdictional claims, we combine our estimate with Synergy Research Group's 2025 estimate that 54% of worldwide compute capacity is physically located within the United States. That compute capacity is therefore clearly subject to U.S. territorial jurisdiction. When combined with our dataset—where U.S.-based companies operate approximately half of foreign infrastructure by investment value—this suggests that the U.S. may exert jurisdictional influence over a substantial share of global compute capacity. This results in an approximate jurisdictional reach of 76% globally (0.54 + (0.46 × 0.48)), or up to 80% for AI data centers (0.54 + (0.46 × 0.56)).

# 5 Implications

## 5.1 Implications for U.S. policy

The potential jurisdictional reach of U.S.-based data center operators may represent the second most far-reaching regulatory lever available to the U.S. for influencing international AI development—second to export controls on key semiconductor technologies. However, the long-term effectiveness of export controls depends on U.S. dominance in key parts of the semiconductor supply chain. Over time, foreign actors may adopt substitute technologies or reconfigure supply chains to reduce exposure to U.S. rules (Lalwani 2025). Furthermore, export controls are in tension with other strategic goals, such as selling U.S. products to global markets, and companies producing technologies subject to export controls can see diminished revenues.

Given these challenges of export controls, jurisdictional claims through U.S. operators may become an even more important governance tool over time. These claims would likely need to be negotiated with the countries hosting the data centers through diplomatic channels, as seen in post–CLOUD Act negotiations on data centers with multiple jurisdictional claims. To preserve or expand the influence of U.S. operators, U.S. policymakers could explore ways to encourage firms to retain operational roles in strategic locations. This topic warrants further research, which our dataset's qualitative entries could offer further insight into.

Still, jurisdictional claims through operators do not grant the U.S. hooks into the full landscape of non-U.S. data centers. For the remaining 52% of data center value that does not use U.S. operators, there are limited regulatory backups to export controls. In particular, we identified significant infrastructure investments in countries such as India, Taiwan, South Korea, and South Africa that—as far as we know—fall outside the reach of both U.S. and Chinese operators. If neither export controls nor operator-based claims applied to their data centers, U.S. influence over these data centers may depend on sustained bilateral or multilateral efforts to align on AI policy.

## 5.2 Implications for policy outside of the U.S.

Each country pursues distinct goals and faces constraints in pursuing those goals, including digital sovereignty. In some cases, maximizing jurisdictional control over data centers may conflict with other objectives—such as deploying high-performance infrastructure for AI applications or attracting private sector investment. Countries should assess their priorities and determine under what conditions foreign-operated data centers may be acceptable or even advantageous. Our findings highlight two major considerations that can inform these assessments: (1) the tradeoff between jurisdictional control and infrastructure performance, and (2) the likelihood that localized projects will succeed in practice.

The first key consideration is the tradeoff between establishing local, high-performance data centers—particularly for AI applications—and minimizing exposure to foreign jurisdictional claims. Our dataset shows that U.S. companies disproportionately operate the investment value of non-U.S. data centers, particularly those associated with AI applications. In some cases, the benefits of having a functional, high-capacity, or AI-optimized data center may outweigh the chance of exposure to foreign jurisdictional hooks through foreign operators.

A second consideration is the frequency of setbacks in localized data center projects. Countries should assess whether they have a strong track record of developing data centers within their country that allow them to reach their goals—or whether persistent barriers, such as regulatory delays or limited market demand, tend to derail them. Some of these barriers may be addressable through policy changes, while others—like geographic constraints or exposure to natural disasters—may be outside a government's control.

Governments that have invested in localized data centers under the banner of digital sovereignty should also consider the conditions under which working with foreign operators may be the most practical or effective option. In cases where governments must work with foreign operators for one reason or another, it may be useful to realistically assess how much resourcing and political capital the country would be willing—or able—to devote to negotiating jurisdictional claims to the data centers should conflicting jurisdictional claims arise.

### 5.3 Limitations

Several limitations qualify the findings presented in this paper. First and foremost, the quality and quantity of the dataset are limited by the availability of publicly reported information at the time of data collection. More detail is available in our first limitations section, found in our Methods section.

Second, 36% of total investment value in our dataset is not attributed to any operator. This gap may be partly explained by the inclusion of planned data centers in our dataset, which do not always have publicly announced operators at the time of reporting. Regardless of the cause, this represents a significant availability gap in the data. This gap implies that our estimates about U.S. operators may underestimate the role of U.S. operators in existing non-U.S. data centers; among the subset of investment for which operator information is available, U.S. firms account for approximately 75% of the value. Even then, the investment value of many projects was unknown.

Third, the dataset offers a partial view of global data center infrastructure and should be interpreted as a sample rather than a comprehensive census. All estimates are rough approximations intended to capture directional trends. Some observed patterns may reflect characteristics of the dataset itself—such as regional reporting biases—rather than actual global distributions. For this reason, we emphasize contextual metrics such as investment value over raw project counts.

Fourth, our frequency analysis treats each project entry as a discrete data center, even though some entries may refer to multi-facility campuses. As a result, simple counts of actor appearances may understate the scale of involvement by certain firms. For example, one project tagged with a U.S. IaaS provider may represent a large, multi-site investment, while another may reflect a smaller, single-site build. This underscores the importance of analyzing actor presence through investment-weighted metrics rather than frequency alone.

### 5.4 Future research

Future research could improve our estimates, interpret our data, and apply our findings to policy decisions.

First, future work could test and refine key assumptions underlying our estimates, for example, that investment value serves as a reasonable proxy for compute capacity. Second, researchers could gather new data to examine whether any patterns have shifted since our data collection concluded in Q3 2024, particularly in light of recent geopolitical developments under the Trump administration. Third, further analysis could improve estimate precision by investigating the projects for which no operator was identified. Follow-up research could examine whether these projects are typically awarded to domestic operators, U.S.-based firms, or other entities.

Finally, the qualitative data in our dataset, including over 1,000 quotes from public sources, could be analyzed to better understand how governments frame their digital sovereignty strategies and infrastructure priorities. These insights may help inform future negotiations around cross-border jurisdictional claims and data center governance.

# 7 Appendices

## 7.1 Appendix 1: Figure 1

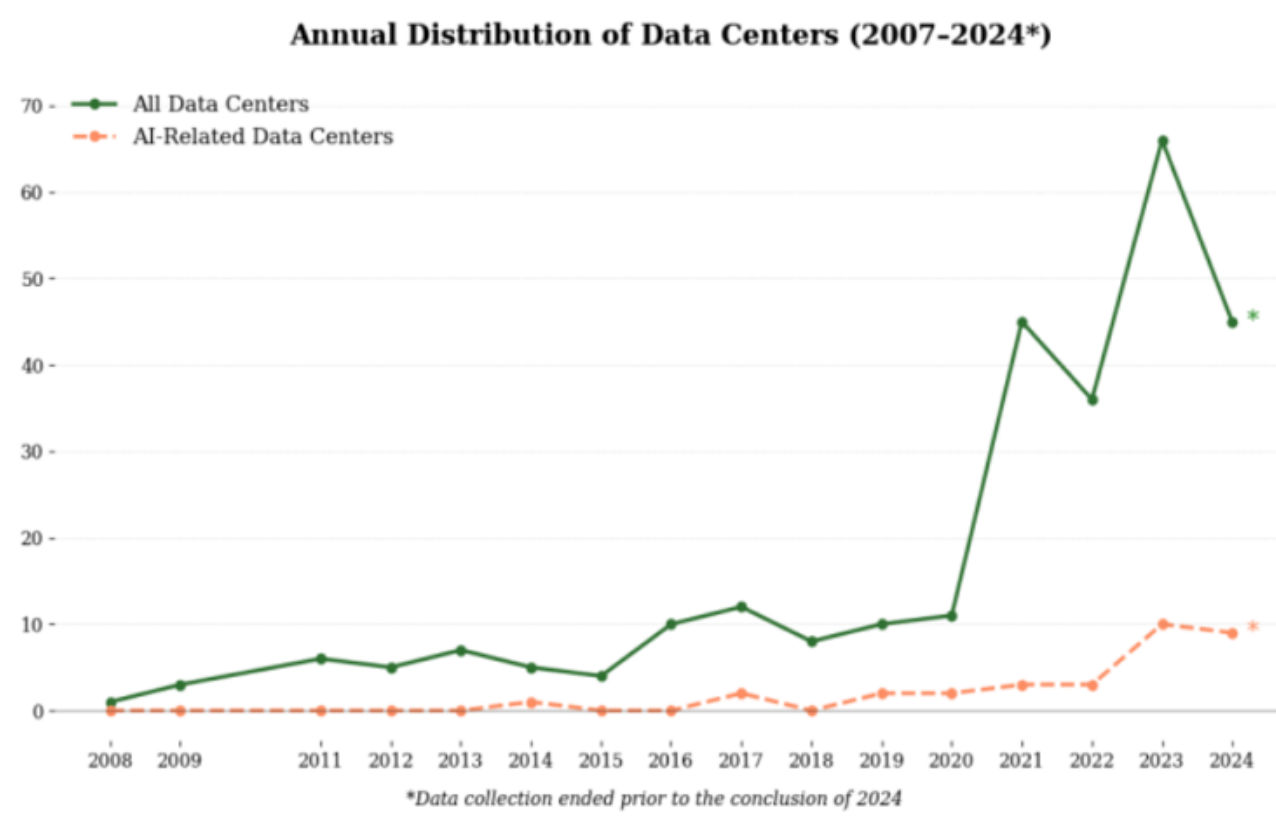

Figure 1: The number of data center projects in our data base announced each year between 2008 and up to quarter 3 of 2024. All data center projects are illustrated in green and a subset of projects that are AI-specific projects are illustrated in orange.

## 7.2 Appendix 2: Table 1

Table 1: Operators of data centers outside of the U.S., rounded to nearest percent.

| Operator Nationality | Frequency | % of Total Investment | Frequency - AI Projects | % of AI Investment |
|---|---|---|---|---|
| U.S. | 18% | 48% | 27% | 56% |
| Chinese | 8% | 5% | 6% | 0% |
| *Chinese firms (outside China)* | 6% | 1% | 1% | 0% |
| Total not attributed to U.S./China | 74% | 47% | 65 % | 44% |
| *Domestic Operator* | 10% | 11% | 16% | 19% |
| *Other Foreign Operator* | 12% | 0% | 1% | 0% |
| *Unattributed* | 53% | 36% | 48% | 25% |

### 7.3 Appendix 3: Figure 2

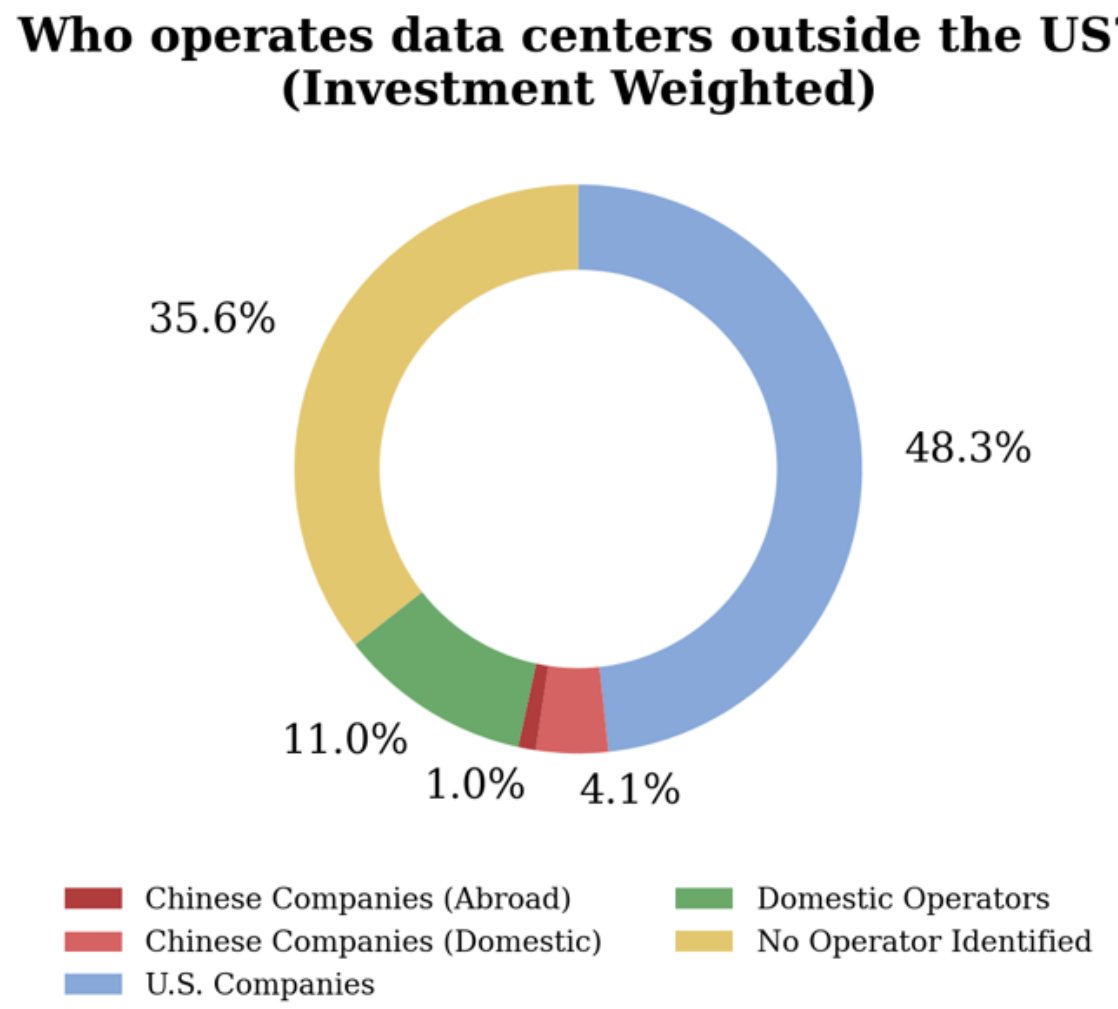

Figure 2: Donut chart showing the distribution of operators of data centers by country.

### 7.4 Appendix 3: Figure 3

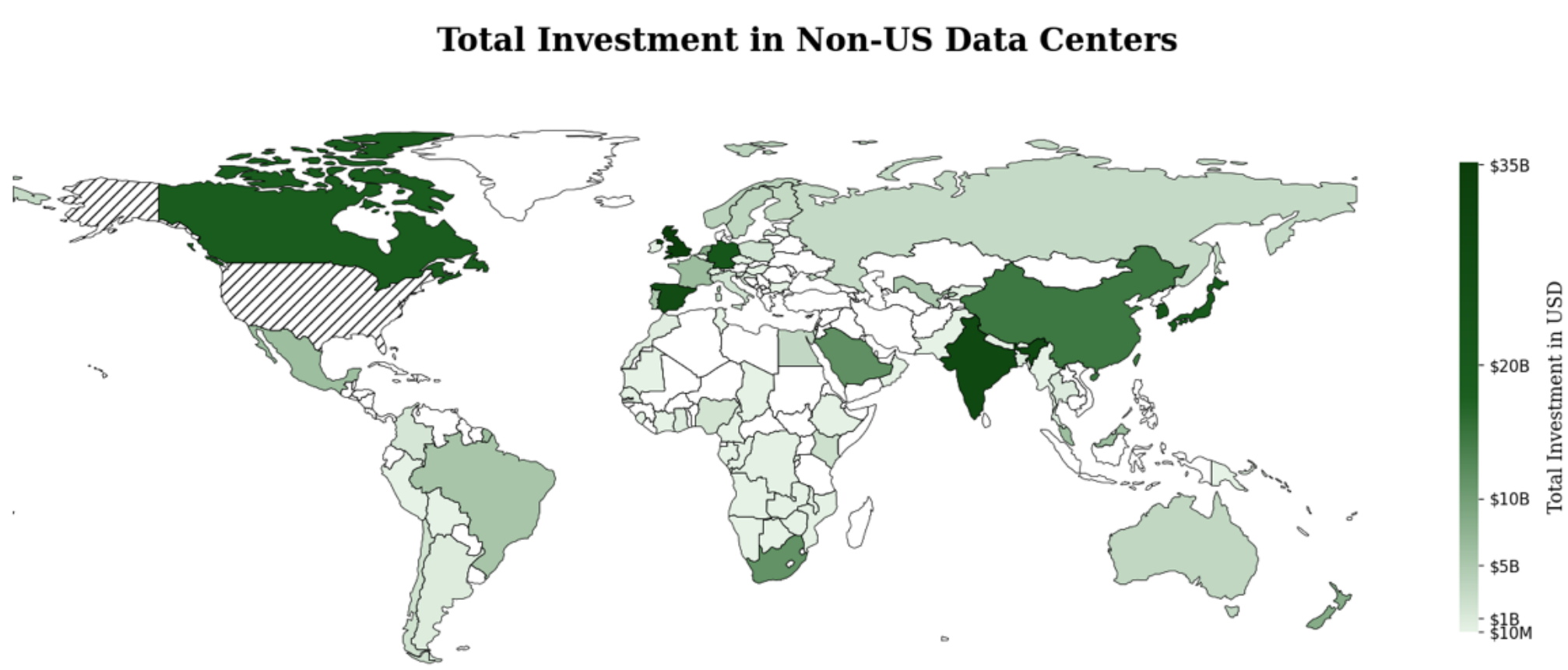

Figure 3: Map showing data center investments (existing and planned) within our dataset by country, not including the U.S. Since our data has limited coverage, these numbers are lower bounds. In total, 360 billion USD were recorded in the dataset.

### 7.5 Appendix 4: Figure 4

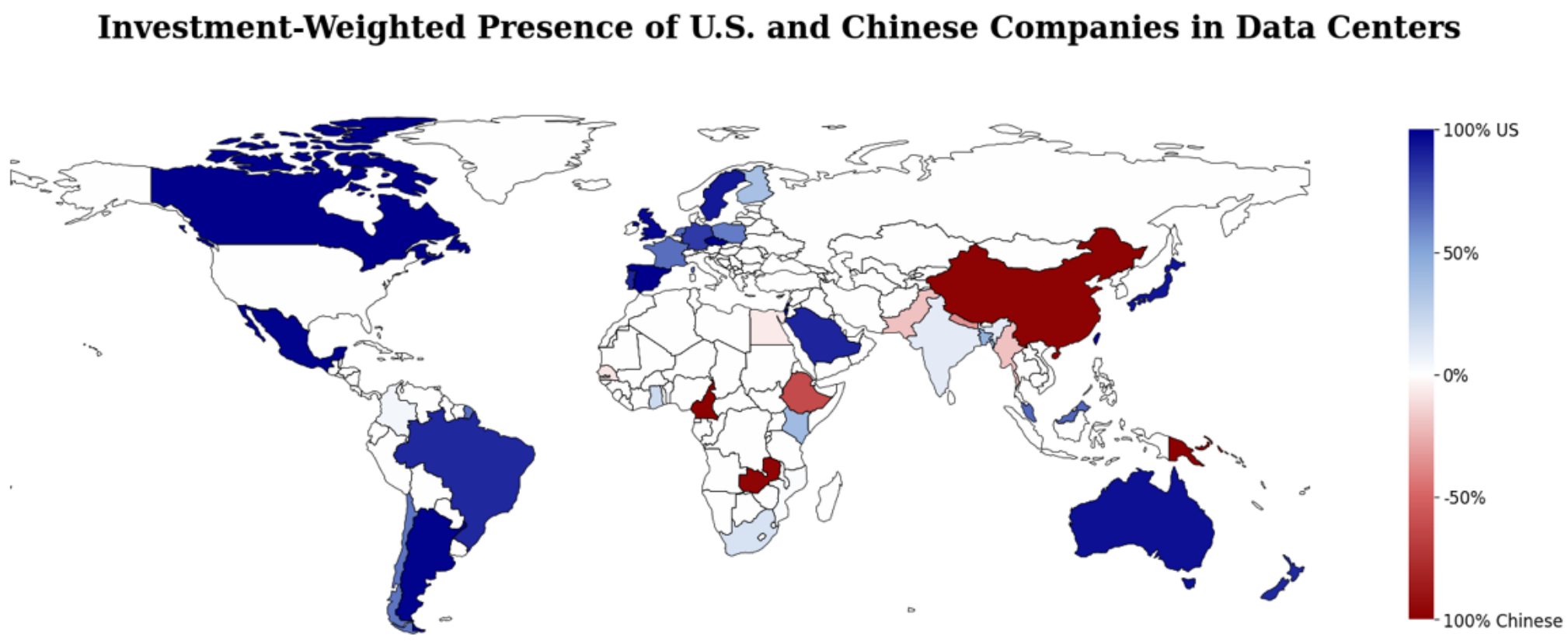

Figure 4: Map showing the relative alignment of countries' data center investments with U.S. or Chinese operators, by investment value. A country appears solid blue if 100% of its investment value is operated by U.S. firms, and solid red if 100% operated by Chinese firms.

### 7.6 Appendix 5: Figure 5

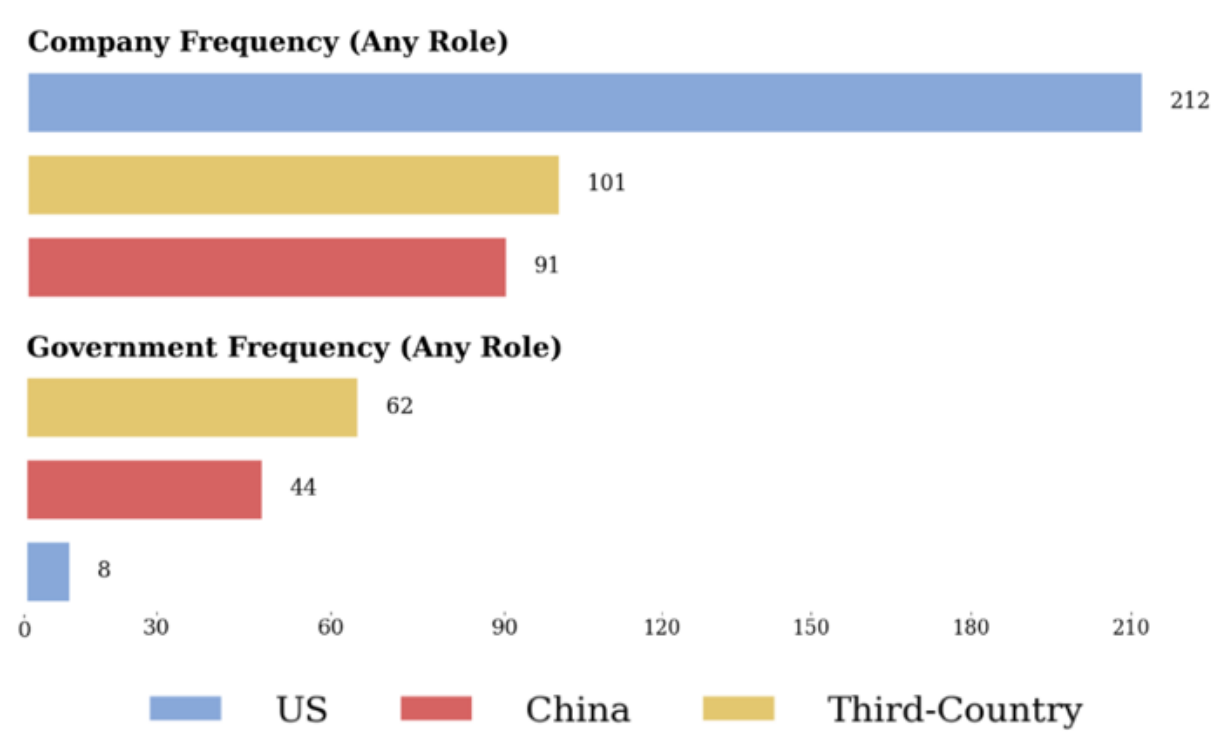

Figure 5: A comparison of the frequency that companies and government entities foreign to that country were recorded in data centers in that country.

### 7.7 Appendix 6: Table 2

Table 2: Countries where at least three different reports on data center projects cited setbacks, sorted by the percent of projects in the country that encountered setbacks. Each country included in the table has over 10 total projects in our dataset.

| Country | Projects with reported setbacks |
|---|---|
| Russia | 36% |
| Malaysia | 30% |
| Pakistan | 30% |
| Ireland | 27% |
| Netherlands | 25% |
| France | 25% |
| Nigeria | 21% |
| Japan | 18% |
| Kenya | 16% |
| China | 14% |
| India | 10% |